# Program Synthesis from Axiomatic Proof of Correctness

Charlie Volkstorf 1/6/15
shymathguy99@aol.com

*ABSTRACT SUMMARY by Section*

## I. Program Synthesis: Specifications to Programs

Program Synthesis is the mapping of a specification of what a computer program is supposed to do, into a computer program that does what the specification says to do. This is equivalent to constructing any computer program and a sound proof that it meets the given specification.

We axiomatically prove statements of the form: program PROG meets specification SPEC. We derive 7 axioms from the definition of the PHP **programming language** in which the programs are to be written. For each primitive function or process described, we write a program that uses only that feature (function or process), and we have an **axiom** that this program meets the specification described. Generic ways to alter or combine programs, that meet known specifications, into new programs that meet known specifications, are our 7 **rules** of inference.

To efficiently prove statements that some program meets a given specification, we work backwards from the specification. We apply the inverses of the rules to the specifications that we must meet, until we reach axioms that are combined by these rules to prove that a particular program meets the given specification. Due to their distinct nature, typically few inverse rules apply. To avoid complex wff and program manipulation algorithms, we advocate the use of simple table maintenance and look-up functions to simulate these complexities as a prototype.

## II. Number Theory in PHP

We are interested in programs written in the PHP programming language that calculate number theoretic functions on positive integers. For example, we want to determine if one number is a factor of another, generate the prime factors of a given number, and list all prime numbers between two given numbers.

Each specification requires that the program decide or list some particular finite set. We represent the set as a Predicate Calculus wff that expresses it, with unquantified variables I, J, K indicating input that is decided, and unquantified variables x, y, z indicating values to list. For example, we define relation LT(a,b) as a<b so that LT(I,J) expresses "Is one given number less than another?" and LT(x,I) expresses "List all numbers less than a given number."

## III. The PHP programming language

We give the commands and functions of a subset of PHP. The echo command indicates output.

IV. Axioms of PHP

We give 7 axioms of PHP, each stating that a particular PHP program meets a particular specification. For example, program "echo $i<$j;" meets specification LT(I,J) with an output of TRUE or FALSE that indicates whether or not input I is less than input J.

V. Rules of Inference for PHP

We give 7 ways that PHP programs that meet known specifications can be altered or combined to create new programs that meet known specifications. For example, a program that decides a set P(I) can be altered to form a program that meets specification ~P(I) to decide the complement of set P. We negate any value that is about to be output.

VI. Example Synthesis 1: BETW(I,J,K) Is one number exclusively between two others?

We give a proof that a particular program decides if one given number is exclusively between two other given numbers. We then show how to discover this program by applying the inverses of the rules to the specification to reach axioms that can be combined by these rules to form the statement that this program meets the given specification.

VII-IX. Synthesis Examples 2-4

| | | |
|---|---|---|
| VII. | 2. BETW(I,x,J) | List all numbers between two numbers. |
| VIII. | 3. (LT(I,J)^EQ(I,x)) v (~LT(I,J)^EQ(J,x)) | Output the minimum of two numbers. |
| IX. | 4. FAC(I,J) | Is one number a factor of another? |

"$B=FALSE ; for ($a=1;!($j<$a);++$a){ $A=FALSE ; if (($a*$i)==$j) $A=TRUE ; if ($A) $B=TRUE ; } ; echo $B ;" and "echo ($j % $i) == 0" # FAC(I,J).

X. Examples 5-10: Listing and Deciding factors, proper factors and prime numbers as Theorems

We summarize how a variety of programs can be easily synthesized by reusing existing proofs.

| | |
|---|---|
| 5. FAC(x,I) | List the factors of a given number. |
| 6. PFAC(x,I) | List the proper factors of a given number. |
| 7. PRIME(I) | Check if a number is prime. |
| 8. FAC(x,I) ^ PRIME(x) | List the prime factors of a given number. |
| 9. PRIME(x) ^ BETW(I,x,J) | List the primes between two given numbers. |
| 10. PRIME(x) ^ BETW("1",x,"100") | List the prime numbers between 1 and 100. |

XI. "for ($a=1 ; !($i<$a) ;++$a) {if (($i%$a) == 0) echo $a ; }" # FAC(x,I) List the factors of I.

XII. "$A=FALSE ; for ($a=1;$a<$i;++$a){ if (1<$a) { if (($i % $a) == 0) $A=TRUE ; } ; } ; echo (!($A)) && (!($i<2)) ;" # PRIME(I) Is I a prime number?

XIII. Additional Useful or Interesting Specifications: We give 10 additional specifications.

XIV. Conclusions and Future Research: We discuss the nature of Program Synthesis using this approach and ways to extend it, as well as its relationship to theoretical computer science.

Program Synthesis from Axiomatic Proof of Correctness

Charlie Volkstorf 1/6/15
shymathguy99@aol.com

*DETAILS*

I. Program Synthesis: Specifications to Programs

PROGRAM SYNTHESIS Program Synthesis is a process in which you input specifications describing what a computer program is supposed to do, and a system outputs a computer program that does what the specifications say, if such a program exists.

PROOF OF CORRECTNESS The only requirement on the program the system produces is that it meet the specification. Thus, having a proof (in a sound system of logic) that the program presented meets the given specification, is the necessary and sufficient condition for this to be Program Synthesis.

AXIOMATIC PROOF OF CORRECTNESS Axiomatic systems of proof use a set of simple known true statements (the axioms) concerning the subject of interest (e.g. natural numbers or sets), and functions (rules of inference) to map such statements into other true statements. Our subject of interest is programs meeting specifications. The definition of the PROGRAMMING LANGUAGE being used provides the simple but encompassing set of true statements for AXIOMS concerning programs meeting specifications. Various ways to alter or combine known programs provide RULES to create additional programs and the specifications they meet as follows:

1. AXIOM: Axioms are statements that certain programs meet certain specifications. These programs are used when we need a program that meets that specification.

2. UNION: To list the union of two sets, we can run programs to list each set.

3. VARIABLE NAMES: We can indicate what is input into a program in a specification and substitute for the corresponding variable names in that program.

4. LOGICAL FUNCTIONS: We create a program that decides a logical function of sets (using Propositional Calculus operators not, and, or) by applying PHP functions that calculate it, to the values output by programs that decide each set.

5. COMBINING PROGRAMS: We combine a program that lists a set with a program that decides a second set, to list the intersection of the two sets. We modify the first program to decide that each element in the first set is also in the second set before it outputs the value.

6. CHECK EXISTS: We can determine if a given set has any elements by modifying a program that lists it, to instead remember that an element exists when it is about to output it.

MEETING THE GIVEN SPECIFICATION To efficiently create programs that meet a particular specification, we work backwards from the given specification by applying the inverses of the rules to create specifications whose programs are mapped by the rules to a statement that some program meets the given specification.

II. Number Theory in PHP

SPECIFICATIONS FOR NUMBER THEORY Our interest is in computer programs that perform functions of Number Theory in the PHP programming language. Our universal set is the positive integers. Examples of useful specifications include:

1. Is a given numbers exclusively between two other numbers?
2. List all numbers exclusively between two other numbers.
3. Output the minimum of two numbers.
4. Is one given number a factor of another?
5. List the factors of a given number.
6. List the prime numbers between 1 and 100.

In each case we are either deciding if a given number is in a given set (1, 4) or listing the numbers in a given set (2, 5, 6), or evaluating a function (3) which is a special case of listing a set. So we either DECIDE or LIST a set. We express the set of numbers using Predicate Calculus. We indicate if each component is input being decided, or output being listed, by its variable name:

Input Variables I, J, K, I4, I5, . . . are values that are input into a wff.
Output Variables x, y, z, x4, x5, . . . are values that are output (listed) from the set expressed.
Quantified Variables A, B, C, A4, A5, . . . appear in e.g. (exists A) and (forall A) in wffs.

In our discussion:

Relation Variables P, Q, R, P4, P5, . . . range over relations over our universal set.
Component Variables a, b, c, a4, a5, . . . represent the components in a relation.

Then P(a) means set P, P(I) means to decide if a given value I is in set P, P(x) means to list the elements of set P, and P(I,x) means for a given input value I to list the set { x | P(I,x) is true. }. We will assume here that all programs must halt, so that all sets listed are finite and after listing a set the program halts.

Altogether, we will be interested in the following 8 relations:

| | | |
|---|---|---|
| 1. Equality | EQ(a,b) | a=b |
| 2. Less Than | LT(a,b) | a<b |
| 3. Between | BETW(a,b,c) | a<b<c |
| 4. Multiply | MUL(a,b,c) | (a*b)=c |
| 5. Factor | FAC(a,b) | (A*a)=b for some A |
| 6. Remainder | REM(a,b,c) | Remainder to a divided by b. e.g. 17 / 3 = 5 remainder 2 |
| 7. Proper Factor | PFAC(a,b) | a is a factor of b and 1<a<b |
| 8. Prime Number | PRIME(a) | a has no proper factors and is not less than 2 |

Then the 6 specifications are:

| | |
|---|---|
| 1. BETW(I,J,K) | Is a given numbers exclusively between two other numbers? |
| 2. BETW(I,x,J) | List all numbers exclusively between two given numbers. |
| 3. (LT(I,J)^EQ(I,x)) v (~LT(I,J)^EQ(J,x)) | Output the minimum of two given numbers. |
| 4. FAC(I,J) | Is one given number a factor of another? |
| 5. FAC(x,I) | List the factors of a given number. |
| 6. PRIME(x) ^ BETW("1",x,"100") | List the prime numbers between 1 and 100. |

We also utilize definitions DEF and their names, that express the equivalence of two wffs using component variables as free variables, to replace a specification with an equivalent one:

| | | |
|---|---|---|
| 1. | BETW | BETW(a,b) = LT(a,b)^LT(b,c) |
| 2. | FAC | FAC(a,b) = (exists A)MUL(A,a,b) |
| 3. | PFAC | PFAC(a,b) = FAC(a,b) ^ BETW("1",a,b) |
| 4. | PRIME | PRIME(a) = ~(exists A)PFAC(A,a) ^ ~LT(a,"2") |
| 5. | REM | FAC(a,b) = REM(b,a,"0") |
| 6. | MUL | MUL(a,b,c) = MUL(b,a,c) |
| 7. | MULT | MUL(a,b,c) = MUL(a,b,c) ^ ~LT(c,a) |
| 8. | ^ | P^Q = Q^P |
| 9. | EQ | P(a) = (exists A)P(A)^EQ(A,a) |

Numbers 1, 2, 3 and 4 simply define relations BETW, FAC, PFAC and PRIME in terms of other relations, and are added for convenience. Number 5 REM is a connection between factors and division with remainder. A remainder of zero defines a factor. Number 6 is commutativity of multiplication. Number 7 MULT is a property of multiplication due to our use of only positive integers. Since for all B we have B > 1, then A*B > A*1 i.e.. A*B > A. If MUL(a,b,c) then a*b>a. Since c = a*b we have c > a i.e. ~(c < a) so MUL(a,b,c) => ~LT(c,a). Note that for any P and Q, if P=> Q then the DEF is P = P^Q. In number 7 MULT, P = MUL(a,b,c) and Q = ~LT(c,a). Number 8 ^ is commutativity of conjunction. Number 9 EQ is a property of equality as follows: If P(a), then a is an A for which P(A) and A=a. Conversely, if A=a then a has any property that A has, including P(A), so P(a).

III. The PHP programming language

A PHP program consists of a sequence of commands that operate on expressions that return a value. An expression can contain:

1. Boolean Literals: TRUE , FALSE
2. Numeric Literals: 1 , 2 , 3 , . . .
3. Variables: $ followed by an alphanumeric string
4. Unary Logical Operator: ! (negation)
5. Binary Logical Operators: && (and) , | | (or) , = = (equality) , < (less that)
6. Binary Numeric Operator: * (multiplication) , % (remainder)
7. Parentheses to treat an expression as a single value: ( )

A program is a series of commands of the following types, which use expressions E1, E2 and E3. Their use is axiomatized by the axioms and rules listed after each command.

1. echo E1 ;
   Output the current value of expression E1. (Axiom 1, Rules QUIT, NOT, AND, DO, IF)

2. $A = E1 ;
   Assign the current value of expression E1 to variable $A or any other variable. (Rule QUIT, Axioms 6, 7)

3. if (E1) COMMAND
   Skip COMMAND if E1 is FALSE. (Rules IF and DO)

4. for (E1 ; E2 ; E3) COMMAND
   Execute E1 and then as long as E2 equals TRUE evaluate E2, then execute COMMAND, then execute E3. (Axioms 5, 7)

5. ++$A ;
   Add 1 to variable $A or any other variable. (Axioms 6, 7)

6. {} Grouping of commands
   { Program } is treated as a single COMMAND. (Rule DO, IF, QUIT)

## IV. Axioms of PHP

The expression PROG # SPEC will mean that PHP program PROG meets specification SPEC. We will axiomatize and prove statements of the form PROG # SPEC including a statement in which the SPEC is equal to a particular value input by the user or created by the system as an intermediate goal to proving a statement of that form. By convention we will use the following variable names in programs occurring in axioms:

Input Variables $i , $j , $k , $i4, $i5 , . . . are input into a program and are never assigned a value.
Program Variables $a , $b , $c , $a4 , $a5 , . . . are variables assigned a value.

The rank of a variable name is its position number in the sequence, with the first variable name having a rank of 1 so e.g. $c has rank 3.

For increased legibility, we can also substitute user defined names for Input and Program Variables based on the context in which they occur. A name can be associated with each component of each relation in which the variable occurs in the specification.

Then we have the following 7 axioms of PHP:

1. Identity Function: "echo $i ;" # EQ(I,x)
   The echo command outputs the value equal to the input.

2. Equality: "echo $i==$j" # EQ(I,J)
   The == operator decides if two numbers are equal.

3. Decide Less Than: "echo $i < $j ;" # LT(I,J)
   The < operator decides if one number is less than another.

4. Multiplication: "echo $i*$j;" # MUL(I,J,x)
   The * operator multiplies its two arguments.

5. Remainder: "echo $i%$j ;" # REM(I,J,x)

6. List numbers less than the input: "for ($a=1 ; $a<$i ; ++$a) echo $a ;" # LT(x,I)
   FOR loops through all positive integers less than a given positive integer.

7. List numbers less than or equal to the input: "for ($a=1 ; !($i<$a) ; ++$a) echo $a ;" # ~LT(I,x)
   FOR loops through all positive integers less than or equal to a given positive integer.

V. Rules of Inference for PHP

We will utilize the following functions when specifying the rules that alter and combine programs in PHP that meet known specifications to create new programs that meet known specifications:

1. s( M , N ) = Program M with program N substituted for each echo command and argument and the rank of each Program Variable in N that is set increased by the maximum rank of any Program Variable in M, so N will assign values to different variables than M.

2. For any program M the expression [M] within the s function means the argument of the current echo command and argument being replaced by the s function.

Combining (1) and (2): For example, suppose M = "if (1<$i) echo $i*$i ;"  Then s(M,"echo [M]*2;") = "if (1<$i) echo $i*$i*2 ;"  since [M] = "$i*$i" the argument of the one echo command and [M]*2 = "$i*$i*2".  The echo command "echo $i*$i;" is replaced with "echo [M]*2;" where [M]="$i*$i" so "echo [M]*2" = "echo $i*$i*2".

3. For any program M and expressions E1, E2, . . ., the expression M:I=E1,J=E2,. . . is program M with each reference to variable I replaced by E1, each reference to J replaced by E2 etc.  For example, if M = "if ($a<$i)||($a<$j)" then M:I=J,J=K = "if ($a<$j)||($a<$k)" .  Any input variable names can be used in place of I and J.

4. A period means string concatenation.

Then the 7 Rules of Inference of PHP are:

1. SUB:I=J   M # P(I) => M:I=J # P(J)

This means that if program M meets specification P(I), then if we substitute the variable for J in place of the variable for I, the modified M meets specification P(J).  For example, if M is "echo $i<3;" then M # LT(I,"3") , M:I=J is "echo $j<3;" , and M:I=J # LT(J,"3") .

2. NOT:  M # P => s( M , "echo !([M]);" ) # ~P

This means that for any program M that meets a specification of the form P (decide a set), the program created by substituting "echo !([M])" for the echo command (where [M] is the argument of that echo command) decides ~P, the complement of P.  For example, "echo $i<$j ;" # LT(I,J) i.e. program "echo $i<$j ;" decides if I<J, so "echo !($i<$j) ;" # ~LT(I,J) i.e. program "echo !($i<$j) ;" decides if ~ (I<J).

3. AND: M # P , N # Q => s( M , s( N , "echo ([M])&&([N]) ; ") ) # P ^ Q

This means if program M decides set P and program N decides set Q, then we can create a program to decide the conjunction P^Q of P and Q as follows: Function s takes program M that decides P, and when M is about to output (echo) the value of P, function s substitutes program N that instead decides set Q. But when program N is about to output the value of Q, function s substitutes "echo ([M])&&([N]) ; " to output the value of ([M])&&([N]), which is P^Q, instead. This is the value of P that program M was about to output, and the value of Q that N was about to output, operated on by && which is the conjunction of P and Q.

4. DO: List P + Decide Q => List P^Q

We can combine a program that lists a set P with a program that decides a set Q to produce a program that lists the intersection P^Q of sets P and Q. Alter the program that lists P to check that each output is also in Q. As Q may have multiple input variables, we indicate which input variable in the program that decides Q receives output x from P(x) at the DO and in references to the program N that decides set Q. We indicate I=X or J=X at the DO, and I=[M] or J=[M] within the program, to show the expression taken from program M and where it is input into program N. Examples:

DO:I=X
M # P(x) , N # Q(I) => s( M , s( N:I=[M] , "{ if ([N:I=[M]]) echo [M] ; } ;" ) ) # P(x) ^ Q(x)

Function s starts with program M that lists set P, and when it is about to output an element of P, function s instead runs program N, that decides set Q, with the output from M substituted for input variable I in program N, to decide if the element of P being output is also in set Q. The PHP if command checks that program N would output TRUE before the output of the element of P from program M.

DO:I=X
M # P(x) , N # Q(I,J) => s( M , s( N:I=[M] , "{ if ([N:I=[M]]) echo [M] ; } ;" ) ) # P(x) ^ Q(x,I)

DO:J=X
M # P(x) , N # Q(I,J) => s( M , s( N:J=[M] , "{ if ([N:J=[M]]) echo [M] ; } ;" ) ) # P(x) ^ Q(I,x)

We can substitute x from P(x) for either I or J in Q(I,J), and indicate which in the DO rule specifications DO:I=X and DO:J=X. Then Q has one less I variable and we rename the I variables to normalize them.

5. IF: M # P , N # Q(x) => s( M , s( N , "{ if ([M]) echo [N] ; } ;" ) ) # P ^ Q(x)

The specification P^Q(x) means that if P is true the program lists set Q, and if P is false then there is no output. So we take a program that is about to output the value of P and instead run a program to list the elements of set Q but instead output each only if P is true. (Alternately, we could skip the program to list Q altogether if P is false or call M from N.)

6. UNION: M # P(x) , N # Q(x) => M . N # P(x) v Q(x)

To list all elements of the union of two sets, we can simply list the elements of one set, then list the elements of the other set. Note that there may be duplicates.

7. QUIT: M # P(x) => $A=FALSE; . { s( M , "$A=TRUE;" ) } . echo $A; # (exists A)P(A)

To determine if there are any elements in set P, we list the elements of P but rather than output the value, we keep $A equal to TRUE or FALSE to indicate whether we would have output any values. If any element is about to be output we instead assign TRUE to $A. Then we output $A.

## VI. Example Synthesis 1: BETW(I,J,K) and Backwards Proofs

BETW(I,J,K) Is a given numbers exclusively between two other numbers?

A proof to create a program to decide BETW(I,J,K) follows. Each line is an axiom, DEF or Rule applied to earlier lines.

```
1. Axiom 3            "echo $i<$j;" # LT(I,J)
2. SUB: I=J, J=K 1    "echo $j<$k;" # LT(J,K)
3. AND 1,2            "echo ($i<$j)&&($j<$k);" # LT(I,J) ^ LT(J,K)
4. DEF-BETW 3         "echo ($i<$j)&&($j<$k);" # BETW(I,J,K)
  qed
```

Line 1 uses axiom 3. Line 2 applies the SUB rule to the axiom in line 1. Line 3 applies the AND rule to lines 1 and 2. Line 4 applies DEF BETW to line 3, producing a statement that some program meets the specification wanted.

The AND Rule states that M # P , N # Q => s( M , s( N , "echo ([M])&&([N]) ; ") ) # P ^ Q .
In line 3, M = "echo $i<$j;" and N = "echo $j<$k;" . Then s( N , "echo ([M])&&([N]) ; ") = "echo ($i<$j)&&($j<$k);" . Also s( M , s( N , "echo ([M])&&([N]) ; ") ) = "echo ($i<$j)&&($j<$k);" since M contains only the echo command.

BACKWARDS PROOFS Then "echo ($i<$j)&&($j<$k);" is a program that meets specification BETW(I,J,K). We can efficiently discover this proof by working backwards from the original specification. We apply the inverses of rules to specifications to produce axioms that can be combined by these rules to prove a statement that a particular program meets the original specification.

We can represent the process of creating this proof backwards using a list of specifications to meet and the axioms and rules that meet them. Each entry in the list gives one of the following:

1. A specification to meet.
2. An axiom whose program we will use.
3. A rule that will be applied to the last one or two programs encountered so far as arguments.

We transform the specification into a series of axioms and rules by repeatedly replacing the first entry that is a specification with one of the following:

1. An equivalent specification using a DEF.
2. An axiom that meets that specification.
3. An axiom and substitution that meets that specification.
4. A rule and specifications as arguments that map to the given specification.
5. A theorem - a program that was synthesized earlier to meet that specification.

A backwards proof that some particular program meets specification BETW(I,J,K) follows. Wffs in **bold** are those created by the current step.

1. We begin with:

   **BETW(I,J,K)**

2. Apply DEF BETW to entry 1 to use relation LT instead of BETW:

   **LT(I,J)^LT(J,K)**

3. Apply inverse of AND rule to entry 1:

   **LT(I,J)**
   **LT(J,K)**
   **AND**

4. Replace entry 1 LT(I,J) with Axiom 3 because Axiom 3 gives a program for LT(I,J).

   **Axiom 3 = LT(I,J)**
   LT(J,K)
   AND

5. Replace entry 2 LT(J,K) with Axiom 3 plus a SUB for:

   Axiom 3 = LT(I,J)
   **Axiom 3 = LT(I,J)**
   **SUB:I=J,J=K**
   AND

We now have all axioms and rules, and no more specifications to meet. This represents an expression whose atomic values are programs and whose operations (functions) are rules. We evaluate this expression by starting with an empty execution stack and processing each entry as follows:

1. Axiom: Add its program to the top of the execution stack.
2. Rule: Apply to its arguments on the stack and replace them with the resulting program.

The program for Axiom 3 is "echo $i<$j;" and the program for BETW(I,J,K) is constructed as follows:

| | | |
|---|---|---|
| 1. Axiom 3: | "echo $i<$j;" | Put the program in Axiom 3 on the stack. |
| 2. Axiom 3: | "echo $i<$j;"<br>"echo $i<$j;" | Put the program in Axiom 3 on the stack. |
| 3. SUB: J, K | "echo $j<$k;"<br>"echo $i<$j;" | Apply SUB:J,K to the top entry of the stack. |
| 4. AND: | "echo ($i<$j)&&($j<$k);" | Apply Rule AND to the top two entries. |

This gives us "echo ($i<$j)&&($j<$k);" as a program that meets BETW(I,J,K), i.e. decides if a given number is strictly between two other given numbers

VII. Example Synthesis 2: BETW(I,x,J)

List all numbers exclusively between two given numbers.

The backwards proof:

| | | |
|---|---|---|
| 1. Given Spec | **BETW(I,x,J)** | List the numbers exclusively between two numbers. |
| 2. DEF-BETW | **LT(I,x) ^ LT(x,J)** | definition of BETW replaces the BETW reference |
| 3. DEF ^ | **LT(x,J) ^ LT(I,x)** | reverse the order of the arguments of ^ |
| 4. DO:K=x | **LT(x,J)**<br>**LT(I,K)**<br>**DO:K=x 1,2** | DO rule replaces the one entry.<br><br>The x in LT(I,x) becomes K to form LT(I,K) |
| 5. AX6 | **LT(x,I) = AX6**<br>**SUB: I=J**<br>LT(I,K)<br>DO:K=x 2,3 | Create LT(x,J) at entry 1 using Axiom 6 and SUB. |
| 6. AX3 | LT(x,I)=AX6<br>SUB: I=J<br>**LT(I,J) =AX3**<br>**SUB: J=K**<br>DO: K=x 2,4 | Create LT(I,K) at entry 3 using Axiom 3 and SUB. |

We now have the specification expressed as axioms and rules applied to them:

LT(x,I)=AX6
SUB: I=J
LT(I,J) =AX3
SUB: J=K
DO:K=x 2,4

Then we create the program by applying each line - axiom or rule:

| | | |
|---|---|---|
| 1. LT(x,I)=AX6 | "for ($a=1;$a<$i;++$a) echo $a;" | Axiom 6 Program for LT(x,I) |
| 2. SUB:I=J | "for ($a=1;$a<$j;++$a) echo $a;" | Change $i to $j to get LT(x,J) |
| 3. LT(I,J) =AX3 | "echo $i<$j;"<br>"for ($a=1;$a<$j;++$a) echo $a;" | Add Axiom 3 Program for LT(I,J) |
| 4. SUB: J=K | "echo $i<$k;"<br>"for ($a=1;$a<$j;++$a) echo $a;" | Change $j to $k to get LT(I,K) |
| 5. DO:K=x 2,4 | "for ($a=1;$a<$j;++$a) { if ($i<$a) echo $a; } ;" | |

DO:K=x:
M # P(x) , N # Q(I) => s( M , s( N:K=[M] , "{ if ([N:K=[M]]) echo [M] ; } ;" ) ) # P(x) ^ Q(x)

M = "for ($a=1;$a<$j;++$a) echo $a;"     P(x) = LT(x,J)
N = "echo $i<$k;"     Q(I) = LT(I,K)

So,
[M] = "$a"     [N] = "$i<$k"
N:K=[M] = "echo $i<$a"     [N:K=[M]] = "$i<$a"
"{ if ([N:K=[M]]) echo [M] ; } ;" = { if ($i<$a) echo $a ; } ;"

So "for ($a=1;$a<$j;++$a) { if ($i<$a) echo $a; } ;" # BETW(I,x,J) i.e. lists the numbers between the two given numbers I and J.

VIII. Example Synthesis 3: (LT(I,J)^EQ(I,x)) v (~LT(I,J)^EQ(J,x)) Minimum of Two Numbers

Output the minimum of two numbers. The backwards proof:

| | | |
|---|---|---|
| 1. User | **(LT(I,J)^EQ(I,x)) v (~LT(I,J)^EQ(J,x))** | Add the user specification to the list. |
| 2. UNION | **LT(I,J)^EQ(I,x)**<br>**~LT(I,J)^EQ(J,x)**<br>**UNION 1,2** | Apply the inverse of the UNION Rule to entry 1. |
| 3. IF | **LT(I,J)**<br>**EQ(I,x)**<br>**IF 1,2**<br>~LT(I,J)^EQ(J,x)<br>UNION 3,4 | Apply the inverse of the IF Rule to entry 1 LT(I,J)^EQ(I,x). |
| 4. AX | **LT(I,J) = AX3**<br>EQ(I,x)<br>IF 1,2<br>~LT(I,J)^EQ(J,x)<br>UNION 3,4 | Use Axiom 3 for entry 1 LT(I,J). |
| 5. AX | LT(I,J) = AX3<br>**EQ(I,x) = AX1**<br>IF 1,2<br>~LT(I,J)^EQ(J,x)<br>UNION 3,4 | Use Axiom 1 for entry 2 EQ(I,x). |
| 6. IF | LT(I,J) = AX3<br>EQ(I,x) = AX1<br>IF 1,2<br>**~LT(I,J)**<br>**EQ(J,x)**<br>**IF 4,5**<br>UNION 3,6 | Apply Rule IF to entry 4 ~LT(I,J)^EQ(J,x). |
| 7. NOT | LT(I,J) = AX3<br>EQ(I,x) = AX1<br>IF 1,2<br>**LT(I,J) = AX3**<br>**NOT 4**<br>EQ(J,x)<br>IF 5,6<br>UNION 3,7 | Apply NOT Rule to entry 4 ~LT(I,J) . |
| 8. AX | LT(I,J) = AX3 | Apply Axiom 1 and SUB to entry 6 EQ(J,x) . |
| | EQ(I,x) = AX1 | |
| | IF 1,2 | |
| | LT(I,J) = AX3 | |
| | NOT 4 | |
| | **EQ(I,x) = AX1** | |
| | **SUB: I=J** | |
| | IF 5,7 | |
| | UNION 3,8 | |

Then the program is synthesized by axioms and rules:

| | | | |
|---|---|---|---|
| 1. | LT(I,J) = AX3 | echo $i<$j ; | Use Axiom 3. |
| 2. | EQ(I,x) = AX1 | echo $i ; | Use Axiom 1. |
| 3. | IF 1,2 | { if ($i<$j) echo $i ; } ; | Apply IF rule to lines 1 and 2. |
| 4. | LT(I,J) = AX3 | echo $i<$j ; | Use Axiom 3 |
| 5. | NOT 4 | echo !($i<$j) ; | Apply NOT rule to line 4 |
| 6. | EQ(I,x) = AX1 | echo $i ; | Use Axiom 1. |
| 7. | SUB:I=J | echo $j ; | Substitute J for I. |
| 8. | IF 5,7 | if (!($i<$j)) echo $j ; | Apply IF rule to lines 5 and 7. |
| 9. | UNION 3,8 | { if ($i<$j) echo $i ; } ; { if (!($i<$j)) echo $j ; } ; | Apply UNION to 3, 8. |

Then "{if ($i<$j) echo $i;} ; {if (!($i<$j)) echo $j;} ; " # (LT(I,J)^EQ(I,x)) v (~LT(I,J)^EQ(J,x)) i.e. outputs the minimum of two given numbers.

IX. Example Synthesis 4. Is one number a factor of another?  FAC(I,J)

We summarize the synthesis of a program to determine if one number is a factor of another.

"Is one number a factor of another?"

1. **FAC(I,J)** User

2. **(eA)MUL(A,I,J)** DEF FAC 1  Use DEF to replace entry 1 with an equivalent wff.

3. **MUL(x,I,J)** QUIT 1
   **QUIT 1**

4. **MUL(x,I,J)^~LT(J,x)** DEF MUL 1
   QUIT 1

5. **~LT(J,x)^MUL(x,I,J)** DEF ^ 1
   QUIT 1

6. **~LT(J,x)** DO 1
   **MUL(K,I,J)**
   **DO:K=x 1,2**
   QUIT 3

7. **~LT(I,x) = AX 7** AX7 , SUB:I=J 1 ~LT(J,x)
   **SUB:I=J 1**
   MUL(K,I,J)
   DO:K 2,3
   QUIT 4

8. ~LT(I,x) = AX 7 DEF EQ 3 MUL(K,I,J)
   SUB:I=J 1
   **(exists A) MUL(K,I,A) ^ EQ(A,J)**
   DO:K=x 2,3
   QUIT 4

9. ~LT(I,x) = AX 7 QUIT 3 (exists A) MUL(K,I,A) ^ EQ(A,J)
   SUB:I=J 1
   **MUL(K,I,x) ^ EQ(x,J)**
   **QUIT 3**
   DO:K=x 2,4
   QUIT 5

10. ~LT( I , x ) = AX 7 DO 3 MUL(K,I,x) ^ EQ(x,J)
SUB:I=J
**MUL( K , I , x )**
**EQ( L , J )**
**DO:L=x 3,4**
QUIT 5
DO:K=x 2,6
QUIT 7

11. ~LT(I,x) = AX 7 AX 4 + SUB:I=K,J=I 3 MUL(K,I,x)
SUB:I=J
**MUL(I,J,x) = AX 4**
**SUB: I=K,J=I**
EQ(L,J)
DO:L=x 3,5
QUIT 6
DO:K=x 2,7
QUIT 8

12. ~LT(I,x) = AX 7 AX 2 + SUB:I=L 5
SUB:I=J
MUL(I,J,x) = AX 4
SUB:I=K,J=I
**EQ(I,J) = AX 2**
**SUB:I=L**
DO:L=x 4,6
QUIT 7
DO:K=x 2,8
QUIT 9

Then the program is defined by:

1. ~LT(I,x) = AX 7 for ($a=1;!($i<$a);++$a) echo $a ; Use Axiom 7.
2. SUB:I=J for ($a=1;!($j<$a);++$a) echo $a ; Apply SUB to line 1.
3. MUL(I,J,x) = AX 4 echo $i*$j ; Use Axiom 4.
4. SUB:I=K,J=I echo $k*$i ; Apply SUB to line 3.
5. EQ(I,J) = AX 2 echo $i==$j ; Use Axiom 2.
6. SUB:I=L echo $l==$j ; Apply SUB to line 5
7. DO: L=x 4,6 if (($k*$i)==$j) echo $k*$i ; Apply DO to 4, 6.
8. QUIT 7 $A=FALSE ; if (($k*$i)==$j) $A=TRUE ; echo $A ; Apply QUIT to 7

9. DO:K=x 2,8 for ($a=1;!($j<$a);++$a) Apply DO to 2, 8
{ $A=FALSE ; if (($a*$i)==$j) $A=TRUE ; if ($A) echo $a ; } ;

10. QUIT 9 $B=FALSE ; for ($a=1;!($j<$a);++$a) { $A=FALSE ; Apply QUIT to 9
if (($a*$i)==$j) $A=TRUE ; if ($A) $B=TRUE ; } ; echo $B ;

Then,

"$B=FALSE ; for ($a=1;!($j<$a);++$a){ $A=FALSE ; if (($a*$i)==$j) $A=TRUE ; if ($A) $B=TRUE ; } ; echo $B ;"

meets FAC(I,J) i.e. decides if one number is a factor of another. Let us examine what this program is doing and why. It must output TRUE or FALSE to indicate if input I is a factor of input J or not. This means there is a number that when multiplied by I equals J. So it looks to see if it can find such a number.

$B is set to whether we have found such a number or not. So initially $B is FALSE. Then we check every number that might equal J when multiplied by I. If such a number were greater than J, then when it is multiplied by I it would be greater than J, not equal. So we need only consider the numbers not greater than J. In the for, $a will equal each number not greater than J.

Then we check to see if any $a when multiplied by I equals J. We check to see if $a times I equals J. We do so by checking every number that $a times I is equal to, and if any of them is equal to J then $a times I is equal to J. There is only one value that $a times I is equal to, and it is equal to $a*$i .

So the program that checks all values of $a*$i checks only the expression $a*$i once. $A is equal to whether we have found such a $a*$i and is set to TRUE if we find one. Then if this program finds $A to be TRUE then $B is TRUE and it will output TRUE when it outputs $B.

Now consider a second proof of a program meeting specification FAC(I,J).

1. **FAC(I,J)** User

2. **REM(J,I,"0")** DEF REM 1 Use DEF to replace entry 1 with an equivalent wff.

3. **(exists A)REM(J,I,A) ^ EQ(A,"0")** DEF EQ 1

4. **REM(J,I,x) ^ EQ(x,"0")** QUIT 1
   **QUIT**

5. **REM(J,I,x)** DO 1
   **EQ(K,"0")**
   **DO:K=x**
   QUIT

6. **REM(I,J,x) = AX5** AX5+SUB 1
   **SUB I=J,J=I**
   EQ(K,"0")
   DO:K=x
   QUIT

7. REM(I,J,x) = AX5 AX2+SUB 3
   SUB I=J,J=I
   **EQ(I,J) = AX2**
   **SUB I=K,J="0"**
   DO:K=x
   QUIT

Then the program is:

| | |
|---|---|
| REM(I,J,x) = AX5 | echo $i % $j ; |
| SUB I=J,J=I | echo $j % $i ; |
| EQ(I,J) = AX2 | echo $i==$j ; |
| SUB I=K,J="0" | echo $k==0 ; |
| DO:K=x | if (($j % $i) == 0) echo $j % $i ; |

M # P(x) , N # Q(I,J) => s( M , s( N:K=[M] , "{ if ([N:K=[M]]) echo [M] ; } ;" ) ) # P(x) ^ Q(I,x)

| | |
|---|---|
| M = "echo $j % $i ;" | N = "echo $k==0 ;" |
| [M] = $j % $i ; | N:K=[M] = "echo ($j % $i) == 0 ;" |

[N:K=[M]] = "($j % $i) == 0 ;"
"{ if ([N:K=[M]]) echo [M] ; } ;" = { if (($j % $i) == 0) echo $j % $i ; } ;

| | |
|---|---|
| QUIT | $A=FALSE ; if (($j % $i) == 0) $A=TRUE ; echo $A ; |
| Code Rule 1 | echo ($j % $i) == 0 ; |

CR1 $A=FALSE ; if P $A=TRUE ; echo $A IS echo P

Then "echo ($j % $i) == 0" ; # FAC(I,J) i.e. decides if I is a factor or J or not.

X. Examples 5-10: Listing and Deciding factors, proper factors and prime numbers as Theorems

5. List the factors of a given number:  FAC(x,I)

By DEF FAC we have (exists A)MUL(A,x,I).  By DEF MUL we have
(exists A)MUL(x,A,I)^~LT(I,x).  Then ~LT(I,x) and x can only be 1 to I.  Then we can go through ~LT(I,x) (Axiom 7) and check each number for being a factor of I (theorem 4.)  Those that are factors are output.

`"for ($a=1 ; !($i<$a) ;++$a) {if (($i%$a) == 0) echo $a ; }"`  (See XI.)

6. List the proper factors of a given number:  PFAC(x,I)

Using DEF PFAC, these are the factors of I that are between 1 and I.  We can list the factors of I (theorem 5) and check each (DO) for being between 1 and I (theorem 1.)

7. Check if a number is prime.  PRIME(I)

By DEF PRIME, to determine if a number is prime we can list its proper factors (theorem 6), so we can check if the number has any proper factors (QUIT Rule) and if it is not less than 2 (Axiom 3 and NOT Rule.)

`"$A=FALSE ; for ($a=1;$a<$i;++$a){ if (1<$a) { if (($i % $a) == 0) $A=TRUE ; } ; } ; echo (!($A)) && (!($i<2)) ;"`  (See XII.)

8. List the prime factors of a given number:  FAC(x,I) ^ PRIME(x)

We list the factors of the given number (theorem 5) and check each (DO Rule) for being a prime number (theorem 7.)

9. List the prime numbers between two given numbers.  PRIME(x) ^ BETW(I,x,J)

We can list the numbers between any two numbers (theorem 2) and check each (DO Rule) for being prime (theorem 7.)

10. List the prime numbers between 1 and 100.  PRIME(x) ^ BETW("1",x,"100")

We can substitute 1 and 100 for inputs in the program to list the prime numbers between two given numbers (theorem 9.)

XI. List the factors of a given number: FAC(x,I) Theorem 5

| | | |
|---|---|---|
| 1. Given Spec | FAC(x,I) | List the factors of a given number. |
| 2. DEF FAC | (exists A)MUL(A,x,I) | x is factor of I iff there is an A such that A*x=I |
| 3. DEF MUL | (exists A)MUL(x,A,I) | |
| 4. DEF MULT | (exists A)MUL(x,A,I)^~LT(I,x) | |
| 5. DEF MUL | (exists A)MUL(A,x,I)^~LT(I,x) | |
| 6. DEF ^ | ~LT(I,x) ^ (exists A) MUL(A,x,I) | |
| 7. DO | ~LT(I,x)<br>(exists A) MUL(A,J,I)<br>DO:J=x | |
| 8. AX | ~LT(I,x) = AX7<br>(exists A) MUL(A,J,I)<br>DO:J=x | |
| 9. DEF FAC 2 | ~LT(I,x) = AX7<br>FAC(J,I)<br>DO:J=x | |
| 10. THM:4 | ~LT(I,x) = AX7<br>**FAC(I,J) = THM 4**<br>**SUB:I=J,J=I**<br>DO:J=x | |

Program:

| | |
|---|---|
| ~LT(I,x) = AX7 | "for ($a=1 ; !($i<$a) ; ++$a) echo $a ;" |
| FAC(I,J) = THM 4 | "echo ($j % $i) == 0" |
| SUB:I=J,J=I | "echo ($i % $j) == 0" |
| DO:J=x | |

M # P(x) , N # Q(I,J) => s( M , s( N:J=[M] , "{ if ([N:J=[M]]) echo [M] ; } ;" ) ) # P(x) ^ Q(I,x)

M = "for ($a=1 ; !($i<$a) ; ++$a) echo $a ;"
N = "echo ($i % $j) == 0"
[M] = "$a" ;
N:J=[M] = "echo ($i % $a) == 0"
[N:J=[M]] = "($i % $a) == 0"
"{ if ([N:J=[M]]) echo [M] ; } ;" = "if (($i%$a) == 0) echo $a ;"
s( M , s( N:J=[M] , "{ if ([N:J=[M]]) echo [M] ; } ;" ) ) =
s("for ($a=1 ; !($i<$a) ; ++$a) echo $a ;",s("echo ($i % $a) == 0" , "if (($i%$a) == 0) echo $a;"))
s("for ($a=1 ; !($i<$a) ; ++$a) echo $a ;" , "if (($i%$a) == 0) echo $a;")

"for ($a=1 ; !($i<$a) ; ++$a) { if (($i%$a) == 0) echo $a ; }" # FAC(x,I)

XII. The Prime Number Checker (Theorem 7)

Thm 2: "for ($a=1;$a<$j;++$a) { if ($i<$a) echo $a; } ;" # BETW(I,x,J)

Thm 4: "echo ($j % $i) == 0 ;" # FAC(I,J)

Backwards proof of PRIME(I):

| | |
|---|---|
| 1. User | **PRIME(I)** |
| 2. DEF PRIME 1 | **~(exists A)PFAC(A,I) ^ ~LT(I,"2")** |
| 3. AND 1 | **~(exists A)PFAC(A,I)**<br>**~LT(I,"2")**<br>**AND** |
| 4. NOT 1 | **(exists A)PFAC(A,I)**<br>**NOT**<br>~LT(I,"2")<br>AND |
| 5. QUIT 1 | **PFAC(X,I)**<br>**QUIT**<br>NOT<br>~LT(I,"2")<br>AND |
| 6. DEF PFAC 1 | **FAC(x,I) ^ BETW("1",x,I)**<br>QUIT<br>NOT<br>~LT(I,"2")<br>AND |
| 7. DEF ^ 1 | **BETW("1",x,I) ^ FAC(x,I)**<br>QUIT<br>NOT<br>~LT(I,"2")<br>AND |
| 8. DO 1 | **BETW("1",x,I)**<br>**FAC(J,I)**<br>**DO:J=x**<br>QUIT<br>NOT<br>~LT(I,"2")<br>AND |

We could also apply the DO rule at step 7 to the original entry 1 at step 6.

9. THM 2 BETW(I,x,J) 1

**BETW(I,x,J) = THM:2**
**SUB:I="1",J=I 1**
FAC(J,I)
DO:J=x
QUIT
NOT
~LT(I,"2")
AND

10. SUB:I=J,J=I 3

BETW(I,x,J) = THM:2
SUB:I="1",J=I 1
**FAC(I,J) = THM:4**
**SUB: I=J,J=I**
DO:J=x
QUIT
NOT
~LT(I,"2")
AND

11. NOT 8

BETW(I,x,J) = THM:2
SUB:I="1",J=I 1
FAC(I,J) = THM:4
SUB: I=J,J=I
DO:J=x
QUIT
NOT
**LT(I,"2")**
**NOT**
AND

12. AX3 = LT(I,J) 8

BETW(I,x,J) = THM:2
SUB:I="1",J=I 1
FAC(I,J) = THM:4
SUB: I=J,J=I
DO:J=x
QUIT
NOT
**LT(I,J) = AX3**
**SUB:J="2"**
**NOT**
AND

The program for PRIME(I) the Prime Number Checker:

```
1 BETW(I,x,J) = THM:2       BETW(I,x,J)
                            for ($a=1;$a<$j;++$a) { if ($i<$a) echo $a; } ;

2 SUB:I="1",J=I 1           BETW("1",x,I)
                            for ($a=1;$a<$i;++$a) { if (1<$a) echo $a; } ;

3. FAC(I,J) = THM:4         FAC(I,J)
                            echo ($j % $i) == 0 ;

4. SUB: I=J,J=I   FAC(J,I)  FAC(J,I)
                            echo ($i % $j) == 0 ;

5. DO:J=x 2,4          BETW("1",x,I)  ^ FAC(x,I)
                       for ($a=1;$a<$i;++$a) { if (1<$a) { if (($i % $a) == 0) echo $a ; } ; } ;
```

```
DO:J=X
M # P(x) , N # Q(I,J) => s( M , s( N:J=[M] , "{ if ([N:J=[M]]) echo [M] ; } ;" ) ) # P(x) ^ Q(I,x)

M = "for ($a=1;$a<$i;++$a) { if (1<$a) echo $a; } ;"
N = "echo ($i % $j) == 0 ;"
[M] = "$a" ;
N:J=[M] = "echo ($i % $a) == 0 ;"
[N:J=[M]] = "($i % $a) == 0 ;"
"{ if ([N:J=[M]]) echo [M] ; } ;" = { if (($i % $a) == 0) echo $a ; } ;

"for ($a=1;$a<$i;++$a) { if (1<$a) { if (($i % $a) == 0) echo $a ; } ; } ;"
```

This lists the proper factors of I viz. BETW("1",x,I)^FAC(x,I) which is PFAC(x,I) Theorem 6.

```
6. QUIT                $A=FALSE ; for ($a=1;$a<$i;++$a)
                       { if (1<$a) { if (($i % $a) == 0) $A=TRUE ; } ; } ; echo $A ;
```

This meets specification (exists A)PFAC(A,I) i.e.: Is there a proper factor of I?

```
7. NOT                 $A=FALSE ; for ($a=1;$a<$i;++$a)
                       { if (1<$a) { if (($i % $a) == 0) $A=TRUE ; } ; } ; echo !($A) ;
```

This meets specification "Is there no proper factor of I?"

```
8. LT(I,J) = AX 3      echo $i<$j ;
9. SUB: J="2"          echo $i<2 ;
10. NOT                echo !($i<2) ;          I is not less than 2.
```

11. AND

```
$A=FALSE ; for ($a=1;$a<$i;++$a)
{ if (1<$a) { if (($i % $a) == 0) $A=TRUE ; } ; } ;
echo (!($A)) && (!($i<2)) ;
```

This meets specification "Is I a prime number?" PRIME(I)

XIII. Additional Useful or Interesting Specifications

1. Decide if a number is a proper factor of another number.
   PFAC(I,J)

2. Smallest proper factor of a number.
   PFAC(x,I)^(all A)~PFAC(A,I)v~LT(A,x)

3. Smallest prime factor of a number.
   PRIME(x)^FAC(x,I)^(all A)~PRIME(A)v~FAC(A,I)v~LT(A,x)

4. What is the next prime number after input I?
   LT(I,x) ^ PRIME(x) ^ (all A) ~LT(I,A) v ~LT(A,x) v ~PRIME(A)

5. Is there a prime number between two given numbers?
   (exists A)PRIME(A)^BETW(I,A,J)

6. List the composite numbers between two given numbers.
   ~PRIME(x)^BETW(I,x,J)

7. List the factors that two given numbers have in common.
   FAC(x,I)^FAC(x,J)

8. Do two given numbers have any common proper factors?
   (exists A)PFAC(A,I)^PFAC(A,J)

9. Do two given numbers have any common prime factors?
   (exists A)FAC(A,I)^FAC(A,J)^PRIME(A)

10. Output the integer square root of a given number, if it exists.
   MUL(x,x,I)

## XIV. Conclusions and Future Research

### CONCLUSIONS

1. Interestingly, we need not ever know the details of how the synthesized programs work, nor must the system be able to understand what particular algorithms are doing or how they work. Program Synthesis is not about meticulously constructing programs one instruction at a time. Rather, it consists of the much simpler task of merely knowing rudimentary rules of how to combine programs, and applying them to "black box" programs whose internals are irrelevant.

2. The axioms and rules of Program Synthesis prove fundamental facts from the Theory of Computation, e.g.:
   a. Every set listed by an axiom is recursively enumerable.
   b. Every set decided by an axiom is recursive.
   c. The NOT rule proves that the complement of every recursive set is recursive.
   d. The AND rule proves that the intersection of two recursive sets is recursive.

3. Recursion Theory uses the same concepts and can use the same formalizations as Program Synthesis. For example, a fundamental theorem from Recursion Theory is that there is a program that outputs itself. That would be a value of M for which M # M or M # EQ(M,x).

### FUTURE RESEARCH

1. Implement Program Synthesis in the target language e.g. PHP.
   a. Use tables for wff and program manipulations as a prototype.
2. Synthesize programs for the 20 specifications.
3. Generic functions that replace obviously inefficient program code with more efficient code.
4. Map between the specification wff and the English statement of the request.
5. Prove the rules of inference and replace them with code generators based on the proofs.
6. Include the sort order of the resulting list as part of the specification.
7. Include the creation of temporary tables and programs that use them.

Generalizing the Axioms and Rules: We give axioms and rules for creating PHP programs. What computer science might say in general is that for any programming language (base of computing), there are recursive functions sub-2, not-1, and-2, or-2, do-2, if-2, union-2 and quit-1 such that for all programs M and N, and all relations P and Q, 1-7 below hold. Is this true?

| | |
|---|---|
| 1. SUB | M # P(I) => sub(M,N) # P(N) |
| 2. NOT | M # P => not(M) # ~P |
| 3. AND | M # P , N # Q => and(M,N) # P^Q or(M,N) # PvQ. |
| 4. DO | M#P(x) , N#Q(I) => do(M,N) # P(x)^Q(x) |
| 5. IF | M#P , N#Q(x) => if(M,N) # P^Q(x) |
| 6. UNION | M#P(x) , N#Q(x) => union(M,N) # P(x) v Q(x) |
| 7. QUIT | M # P(x) => quit(M) # (exists A)P(A) |